\documentclass[prb,aps,twocolumn,showpacs, floatfix,preprintnumbers]{revtex4}

\usepackage[applemac]{inputenc}
\usepackage[T1]{fontenc} 
 \usepackage{amsmath} 
\usepackage{subfigure}
\usepackage{lipsum}
\usepackage{amsfonts} 
\usepackage{amssymb, mathrsfs}
\usepackage{braket}
\usepackage{graphicx} 
\usepackage[usenames]{color} 

\usepackage{bbm}

\DeclareMathOperator\arctanh{arctanh}

\def\beq{\begin{equation}}
\def\eeq{\end{equation}}
\def\bsp{\begin{split}}
\def\esp{\end{split}}
\def\bea{\begin{eqnarray}}
\def\eea{\end{eqnarray}}
\def\ba{\begin{array}}
\def\ea{\end{array}}

\def\lb{\left(}
\def\rb{\right)}

\def\l.{\left.}
\def\r.{\right.}

\def\part{\partial}

\def\ket#1{\mid #1 {\cal{i}}}

 \numberwithin{equation}{section}

\makeatletter

\newcommand{\Rmnum}[1]{\expandafter\@slowromancap\romannumeral #1@}
\makeatother
\begin{document}
\preprint{UdeM-GPP-TH-13-227}
\title{Phase transition between quantum and classical regimes for the escape rate of dimeric molecular nanomagnets in a staggered magnetic field}
\author{S. A.  Owerre}
\email{solomon.akaraka.owerre@umontreal.ca}
\author{M. B. Paranjape} 
\email{paranj@lps.umontreal.ca}
\affiliation{Groupe de physique des particules, D\'epartement de physique,
Universit\'e de Montr\'eal,
C.P. 6128, succ. centre-ville, Montr\'eal, 
Qu\'ebec, Canada, H3C 3J7 }

\begin{abstract}
\section*{Abstract}  
We study the phase transition of the escape rate of exchange-coupled dimer
of single-molecule magnets which are coupled either ferromagnetically or antiferromagnetically in a staggered magnetic field and an easy $z$-axis anisotropy. The Hamiltonian for this system has been used to study dimeric molecular nanomagnet [Mn$_4$]$_2$ which is comprised of two single molecule magnets coupled antiferromagnetically.  We generalize the method of mapping a single-molecule magnetic spin problem onto a quantum-mechanical particle to dimeric molecular nanomagnets. The problem is mapped to a single particle quantum-mechanical
Hamiltonian in terms of the relative coordinate and a coordinate dependent reduced mass. It is shown that the presence of the external staggered magnetic field creates a phase boundary separating the first- from the second-order transition. With the set of parameters used by R. Tiron, $\textit{et al}$ , \prl {\bf 91}, 227203 (2003), and S. Hill, $\textit{et al}$ science  {\bf  302}, 1015 (2003) to fit experimental data for [Mn$_{4}$]$_2$ dimer we find that the critical temperature at the phase boundary is $T^{(c)}_0 =0.29K$. Therefore, thermally activated transitions should occur for temperatures greater than $T^{(c)}_0$.
\end{abstract}


\maketitle


\section{Introduction} 
The study of single-molecule magnets (SMMs) has been the subject of experimental and theoretical interest in recent years. These systems  have been pointed out\cite{cl,chud1} to be a good candidate for investigating  first- and second-order phase transition of the quantum-classical escape rate. The quantum-classical escape rate transition takes place in the presence of a potential barrier, it is mainly in two categories$-$ classical thermal activation over the barrier and quantum tunnelling through the barrier. At high temperatures, transition occurs by classical thermal activation over the barrier while at low-temperatures, transition occurs by quantum tunnelling between two degenerate classical minima. In principle these transitions are greatly influenced by the anisotropy constants and the external magnetic fields. There exits a crossover temperature (first-order transition) $T_0^{(1)}$ from quantum to thermal regime, it is estimated as $T_0^{(1)}=\Delta U/B$, $\Delta U$ is the energy barrier and $B$ is the instanton action responsible for quantum tunnelling. The second-order phase transition occurs for  particles in a cubic or quartic parabolic potential, it takes place at the temperature $T_{0}^{(2)}$, below $T_{0}^{(2)}$  one has the phenomenon of thermally assisted tunnelling and above $T_{0}^{(2)}$ transition occur due to thermal activation to the top of the potential barrier\cite{ cl, solo4,chud1}. 

Garanin and Chudnovsky\cite{cl} have studied the model of a uniaxial single ferromagnetic spin with a transverse magnetic field, which is believed to describe the molecular magnet Mn$_{12}$Ac with a total spin of $s=10$. They showed by using the method of spin-particle mapping \cite{solo,solo1, wznw}, that the phase transition can be understood in analogy of Landau's theory of phase transition, with the free energy expressed as $F = a\psi^2+b\psi^4 +c\psi^6$, where $a=0$ determines the quantum-classical transition and $b=0$ determines the boundary between the first- and second-order phase transition. Many authors \cite{solo2,solo4,solo6,kal} have searched for the possibility of these transitions in the biaxial single ferromagnet spin systems. To the best of our knowledge,  the possibility of these transitions for  exchange-coupled dimer spin systems has not been reported in any literature. In many cases of physical interest, the spins in a physical system, in principle interact with each other either  ferromagnetically or antiferromagnetically.  One physical example in which these interactions occur is the molecular wheels such as Mn$_{12}$\cite{wal, mula,had}, and the molecular dimer [Mn$_{4}$]$_2$\cite{da,aff}, which comprises two Mn$_4$ SMMs of equal spins $s_A=s_B=9/2$, which are coupled antiferromagnetically. These systems are usually modelled with two interacting giant sublattice spins. Additional terms such as easy axis anisotropy, transverse anisotropy and an external magnetic field are usually added to the model Hamiltonian. Therefore, the thermodynamic and low-energy properties of these systems can be studied effectively by two interacting large spin Hamiltonian. Due to recent experiment on molecular Mn$_{12}$ wheel\cite{mula,had} and [Mn$_{4}$]$_2$ dimer\cite{wern4,da,aff,parkma} , such effective Hamiltonian has attracted so much attention. In this paper we will study one form of this effective Hamiltonian. 
 \section{Model Hamiltonian}
Consider the effective Hamiltonian of an  exchange-coupled dimer of SMMs such as [Mn$_{4}$]$_2$ in a staggered magnetic field with an easy $z$-axis anisotropy  
\begin{align}
\hat{H}&=J\hat{\bold{S}}_A \cdot \hat{\bold{S}}_B  - D(\hat{S}_{A,z}^{2} +\hat{S}_{B,z}^{2}) +g\mu_B h (\hat{S}_{A,z}-\hat{S}_{B,z})\label{1}\end{align}
where $J$ is the isotropic Heisenberg exchange interaction, and $J>0  (J<0)$  are antiferromagnetic (ferromagnetic) exchange coupling respectively and $D>|J|>0$ is the easy $z$-axis anisotropy, $h$ is the external magnetic field, $\mu_B$ is the Bohr magneton and $g=2$ is the electron $g$-factor.  The last term indicates that there are staggered magnetic fields $-h$ and $h$ applied to the two sublattices A and B respectively. For the antiferromagnetic coupling, the spins are aligned (classically speaking)  antiparallel along the $z$-axis. The anisotropy and the magnetic field  terms in the Hamiltonian create two classical minima located at $\pm z$-axis, these minima (one being  metastable) are separated by an energy barrier, and any spin configuration  can escape from one minimum to the other either by thermal activation over the barrier or by quantum tunnelling through the barrier. We have omitted a fourth order anisotropy term which is very small compare to the easy-axis term. The spin operators obey the usual commutator relation: $\big[\hat{S}_{j\alpha},\hat{S}_{k\beta}\big]=i\epsilon_{\alpha \beta\gamma}\delta_{jk}\hat{S}_{k\gamma}$ $\left(j,k =A,B; \thinspace\alpha, \beta, \gamma  =x,y,z\right)$. 
The Hilbert space of this system is the tensor product of the two spaces $\mathscr{H}=\mathscr{H}_A \otimes \mathscr{H}_B$ with dim$(\mathscr{H})$= $(2s_A+1)\otimes (2s_B+1)$. The basis of $S_{j}^z$ in this product space is given by  $\ket{s_A, m_A}\otimes\ket{s_B, m_B} \equiv\ket{m_A,m_B}$. The eigenvalue of the diagonal term of the Hamiltonian is simply given by
\bea
\mathcal{E}_d=Jm_Am_B-D(m_A^2+m_B^2) +g\mu_B h(m_A-m_B)
\label{1.1}
\eea
Note that for antiferromagnetic coupling, either $m_A$ or $m_B$ should be replaced  with $-m_A$ or $-m_B$, while for ferromagnetic coupling, Eq.\eqref{1.1} is the exact ground state energy of the quantum Hamiltonian, Eq.\eqref{1}, with the eigenstates $\ket{m_A=s_A,m_B=s_B}$ and $\ket{m_A=-s_A,m_B=-s_B}$, these two states are degenerate for $h=0$ or $s_A=s_B=s$. In principle the spectrum of the Hamiltonian Eq.\eqref{1} for antiferromagnetic spin configuration can be found by exact numerical diagonalization for some compounds \cite{da,had}. Similar  models of this form have been extensively studied by different methods\cite{los,wal,gag,k}. Since the individual $z$-components of the spins do not commute with the Hamiltonian (only the total $z$-component of the spins $\hat{S_z} = \hat{S}_{A,z}+ \hat{S}_{B,z}$ commutes), the two antiferromagnetic classical ground states $|\hskip-1 mm\downarrow, \uparrow\rangle$, and $|\hskip-1 mm\uparrow, \downarrow\rangle$,  where $\ket{\downarrow,\uparrow}\equiv \ket{m_A=-s,m_B=s}$ etc, are not exact eigenstates of Eq.\eqref{1}, in principle there should be an energy splitting between these two states due to tunnelling. We showed\cite{ams} via spin coherent state path integral, for $h=0$   that the degeneracy of the two states $\ket{\uparrow,\downarrow}$ and $\ket{\downarrow, \uparrow}$ are lifted by the transverse exchange interaction $J\neq0$ and the energy splitting is proportional to $\lvert J \rvert^{2s}$ corresponding to $2s^{\text{th}}$ order in perturbation theory in the $J$ term. This result had been obtained by  perturbation theory\cite{kim2,bab1,bab2}.  Thus, the ground and the first excited states become the anti-symmetric and symmetric linear coherent superpositions of these two antiferromagnetic classical ground states\cite{bab}.  The form of the Hamiltonian Eq.\eqref{1} has been used to investigate [Mn$_4$]$_2$ dimer\cite{da,los1,aff} for which $s_A=s_B=s=9/2$, thus there are $(2s+1)^2 \times (2s+1)^2 = 100\times100$ matrices which are sparsely populated giving rise to an exact numerical digonalization of 100 non-zero energy states. The parameters use to fit experimental data for this dimer are $J=0.12K$, $D=0.75K$. At zero magnetic field,  it has been demonstrated by density-functional theory that this simple model can reproduce experimental results in  [Mn$_4$]$_2$ dimer with $D=0.58 K$ and $J=0.27 K$ \cite{parkma}. This model also plays a role in quantum computation for investigating controlled-NOT quantum logic gates\cite{B}.
The purpose of this paper is to map this model to a quantum mechanical particle in an effective potential and investigate the influence of the staggered magnetic field on the first- and second-order phase transition between quantum and classical regimes for the escape rate. We will show that the result of spin coherent state path integral can be recovered from this effective potential mapping. We will focus on the case of antiferromagnetic coupling since the form of Hamiltonian we choose does not possess  any ground state tunnelling for the ferromagnetic case.

 \section{Methodology}
  In the spin-particle formalism, one introduces the spin wave function  using the $S_{iz}, i = 1,2$ eigenstates \cite{solo,solo1,wznw}, and the resulting eigenvalue equation is then transformed to a differential equation, which is further reduced to a Schr\"{o}dinger equation with an effective potential and a constant or coordinate dependent mass. In the present problem the spin wave function  can be written in a more general form as

\beq
\psi =\psi_A\otimes\psi_B =\sum_{\substack {m_A=-s_A \\ m_B=-s_B}}^{s_A,s_B} \mathcal{C}_{m_A,m_B} \mathcal{G}_{m_A,m_B}
\label{2.2}
\eeq
where 
\beq
 \mathcal{G}_{m_A,m_B} = \binom{2s_A}{s_A+m_A}^{-1/2}\binom{2s_B}{s_B+m_B}^{-1/2}\ket{ m_A,m_B}
\eeq
It is noted that either $m_A \rightarrow -m_A$ or $m_B \rightarrow -m_B$ since we are interested in the case of antiferromagnetic spin configuration, however,  as we will see later, one can check  that this replacement does not alter the resulting differential equation.
The action of the spin Hamiltonian Eq.\eqref{1} on the spin wave function  Eq.\eqref{2.2} yields an eigenvalue equation
\begin{equation}
\begin{split}
\hat{H}\psi=&\sum_{\substack {m_A=-s_A \\ m_B=-s_B}}^{s_A,s_B}\mathcal{C}_{m_A,m_B}\Bigg[\frac{J (s_A-m_A)(s_B+m_B)}{2}\mathcal{G}_{m_A+1,m_B-1}\\&   +\frac{J (s_A+m_A)(s_B-m_B)}{2} \mathcal{G}_{m_A-1,m_B+1} \\&+\Bigg(J m_Am_B+g\mu_Bh(m_A-m_B)-D(m_A^2 + m_B^2)\Bigg)\\&\times \mathcal{G}_{m_A,m_B}\Bigg] = \mathcal{E}\psi\end{split}
\label{6}
\end{equation}
which can be written in a more compact form as
\begin{equation}
\begin{split}
 \mathcal{E} \mathcal{C}_{m_A,m_B}=& \left[J m_Am_B-D(m_A^2 + m_B^2)+g\mu_B h(m_A-m_B)\right] \mathcal{C}_{m_A,m_B}\\&+\frac{J(s_A-m_A+1)(s_B+m_B+1)}{2}\mathcal{C}_{m_A-1,m_B+1}   \\&+\frac{J(s_A+m_A+1)(s_B-m_B+1)}{2} \mathcal{C}_{m_A+1,m_B-1}  \end{split}
\label{6}
\end{equation}
where $\mathcal{C}_{-s_i-1} = 0 = \mathcal{C}_{s_i+1}$, etc, $i=A,B$. In order to transform this expression, Eq.\eqref{6}  into a differential equation , we introduce the characteristic function \cite{solo,solo1} for the two particles 
%
\begin{align}
\mathcal{F}(x_1,x_2) =\sum_{\substack {m_A=-s_A \\ m_B=-s_B}}^{s_A,s_B} \mathcal{C}_{m_A,m_B}e^{m_Ax_1}e^{m_Bx_2} 
  \label{2.6}
\end{align}
 It is well-known that when the magnetic field is applied along the hard-axis a topological phase (oscillation of tunnelling splitting) is generated due to an imaginary term arising from the Euclidean action  \cite{solo2,l,k} . In the present problem the magnetic field is along the easy-axis, so we do not expect such effect in this model.  In our representation the characteristic function Eq.\eqref{2.6} is not periodic, but by complexifying the variables $x_1$ and $x_2$ one can see that the function satisfies $\mathcal{F}(x_1+2\pi i,x_2+2\pi i)=e^{2\pi i(s_A+s_B)}\mathcal{F}(x_1,x_2)$. The differential equation for $\mathcal{F}$ yields

\begin{equation}
\begin{split}
&-D\left(\frac{d^2\mathcal{F}}{dx_1^2} +\frac{d^2\mathcal{F}}{dx_2^2}\right) -J \cosh\left(x_1-x_2\right) \frac{d}{dx_1}\left(\frac{d\mathcal{F}}{dx_2}\right)\\& + J\frac{d}{dx_1}\left(\frac{d\mathcal{F}}{dx_2}\right)- \lb g\mu_Bh -Js_A\sinh(x_1-x_2)\rb\frac{d\mathcal{F}}{dx_2}\\&+\lb g\mu_Bh-Js_B\sinh(x_1-x_2)\rb\frac{d\mathcal{F}}{dx_1}\\& +(Js_As_B\cosh(x_1-x_2)-\mathcal{E})\mathcal{F}=0
\label{0} 
\end{split}
\end{equation}
 As one expects from two interacting particles, the hyperbolic functions in Eq.\eqref{0} emerge as functions of the relative coordinate. Proceeding in a similar way to that of classical theory, we introduce the relative and center of mass coordinates as
\beq
r =x_1-x_2,\quad q = \frac{x_1 +x_2}{2}
\eeq
then Eq.\eqref{0}  reduces to a second-order differential equation with variable coefficients  in terms of the relative and center of mass coordinates
\begin{equation}
\begin{split}
&\mathcal{P}_1(r)\frac{d^2\mathcal{F}}{dr^2}  + \mathcal{P}_2(r)\frac{d^2\mathcal{F}}{dq^2}+\mathcal{P}_3(r)\frac{d\mathcal{F}}{dr}+ \mathcal{P}_4(r)\frac{d\mathcal{F}}{dq} \\&+(\mathcal{P}_5(r)- \mathcal{E})\mathcal{F}=0
\label{011} 
\end{split}
\end{equation}
where the $\mathcal{P}_i(r)$ functions are given by

\begin{equation}
\begin{split}
&\mathcal{P}_1(r) =-2\left[D +\frac{J}{2} -\frac{J}{2}\cosh r\right] \\&\mathcal{P}_2(r) =-\frac{1}{2}\left[D -\frac{J}{2} +\frac{J}{2}\cosh r\right] ,\thinspace 
\\& \mathcal{P}_3(r)=(2g\mu_Bh-J(s_A+s_B)\sinh r), \\& \mathcal{P}_4(r)=\frac{J(s_A-s_B)}{2}\sinh r,\thinspace \mathcal{P}_5(r)=Js_As_B\cosh r\label{112}
\end{split}
\end{equation}
 and  $\mathcal{F} = \mathcal{F}(r,q)$. 
 
In general, for $s_A\neq s_B$ the exact solution of Eq.\eqref{011} is unknown. But in most cases of physical interest such as molecular magnets and molecular wheels \cite{wern4}, the two spins are equal. Thus,  it is reasonable to consider a special case of equal spins $s_A=s_B=s$. In this case the expression for $\mathcal{P}_4(r)$ vanishes and the rest of Eq.\eqref{011} can then be simplified by separation of variable, $\mathcal{F}(r,q) = \mathcal{X}(r)\mathcal{Y}(q)$. The $q$ dependence of this function is in fact unity, this can be clearly shown from Eq.\eqref{2.6}. The function $\mathcal{F}(r,q)$ can be written explicitly as
 \bea
 \mathcal{F}(r,q) =\sum_{\substack {m_A=-s \\ m_B=-s}}^{s,s} \mathcal{C}_{m_A,-m_B}e^{\frac{(m_A+m_B)r}{2}}\underbrace {e^{\frac{(m_A-m_B)q}{2}}}_{1}
  \label{dimer55}
 \eea
 where $m_B\to -m_B$ as required for antiferromagnetic configuration.
 Hence
\begin{align}
 \mathcal{F}(r,q) = \mathcal{X}(r)=\sum_{\substack {m_A=-s \\ m_B=-s}}^{s,s} \mathcal{C}_{m_A,-m_B}e^{\frac{(m_A+m_B)r}{2}}  
 \label{dimer56}
 \end{align}
 Thus Eq.\eqref{011} reduces to a function of $r$ alone:
\begin{align}
 \mathcal{P}_1(r)\frac{d^2\mathcal{X}(r)}{dr^2} + \mathcal{P}_3(r)\frac{d\mathcal{X}(r)}{dr} +{(\mathcal{P}_5(r)- \mathcal{E})}\mathcal{X}(r) =0
\label{dimer10}
\end{align}
 It is convenient to write this equation  out explicitly ($r\rightarrow r+i\pi$ for convenience\footnote{This transformation is required in order to avoid a negative mass in Eq.\eqref{2.21}}):
\begin{align}
&-2\lb D +\frac{J}{2} +\frac{J}{2}\cosh r\rb\frac{d^2\mathcal{X}}{dr^2} + 2(g\mu_B h+ J s \sinh r)\frac{d\mathcal{X}}{dr} \nonumber\\& -\lb Js^2 \cosh r  +\mathcal{E}\rb\mathcal{X}=0
\label{18}
\end{align}
One step to obtaining a Schr\"{o}dinger equation is to eliminate the first derivative term in Eq.\eqref{18}. This can be done by the transformation   
 \beq
\Psi(r) =e^{-y(r)}\mathcal{X}(r)
 \label{3.15}
  \eeq
 with $y(r)$ given by 
  
\begin{equation}
y(r)= s\ln(2+\kappa+\kappa\cosh r)+\frac{2\tilde{s}\alpha}{\sqrt{1+\kappa}}\arctanh\bigg[\frac{\tanh\lb\frac{r}{2}\rb}{\sqrt{1+\kappa}}\bigg]
\end{equation}
where $\tilde{s}= (s+ \frac{1}{2})$,  $\kappa = J/D$  and $\alpha = g\mu_Bh/2D\tilde{s}$. 

Notice that $\Psi(r)\rightarrow 0$ as $r\rightarrow \pm \infty$, so it can be regarded as the particle's quantum-mechanical reduced wave function. Plugging Eq.\eqref{3.15} into Eq.\eqref{18} we arrive at the  Schr\"{o}dinger equation:
 \begin{figure}
\centering
\includegraphics[scale=0.35]{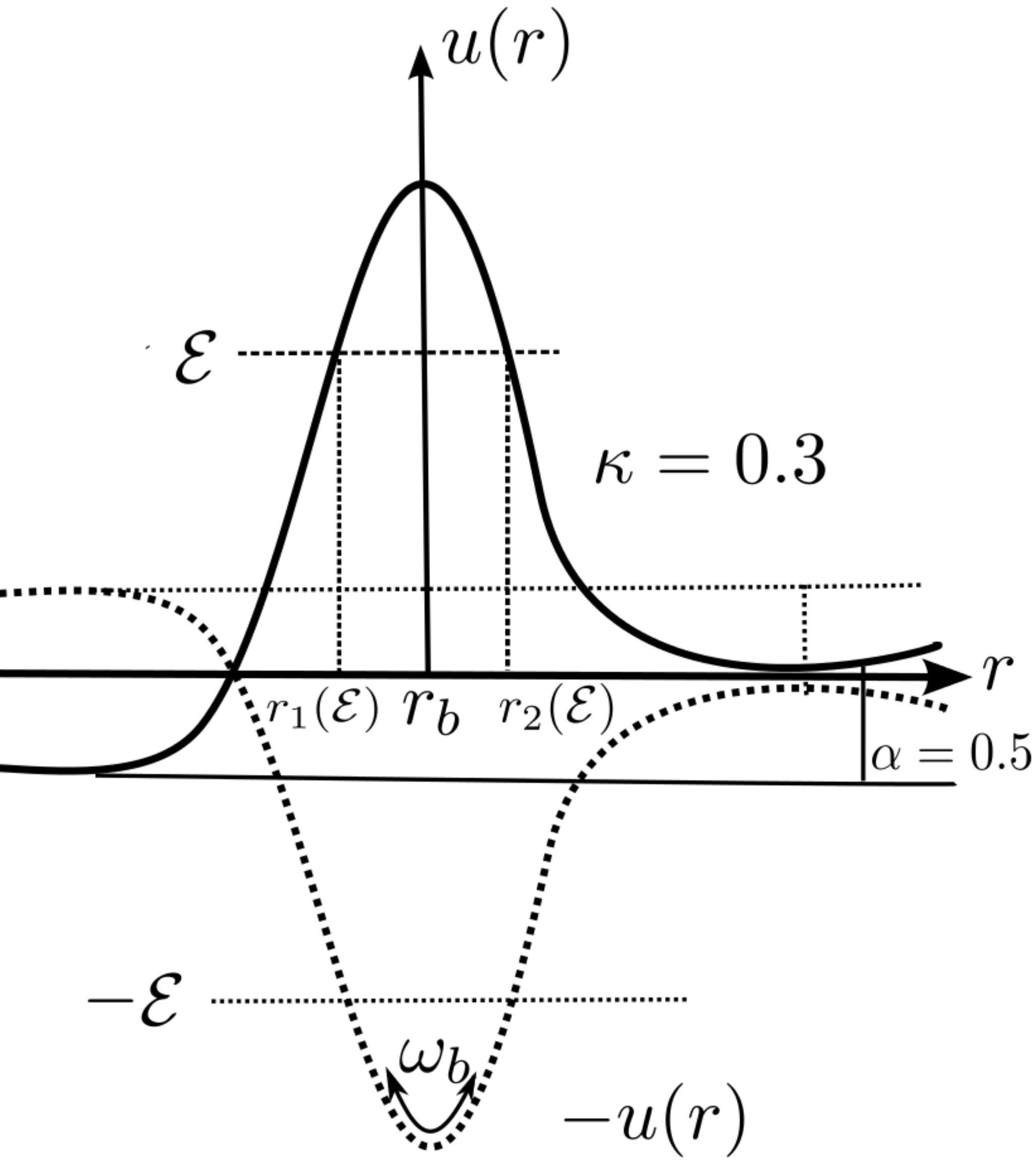}
\caption{The effective potential and its inverse vs. $r$ for $\kappa=0.3$ and $\alpha=0.5$}
\label{interact}
\end{figure}
 
 \beq
 H\Psi(r) = \mathcal{E}\Psi(r)
\label{20}
\eeq
with
\beq
H = -\frac{\nabla^2}{2\mu(r)}  + U(r) , \quad \nabla =\frac{d}{dr}
\label{3.18}
\eeq
The effective potential $U(r) = 2D\tilde{s}^2u(r)$ and the coordinate dependent reduced mass $\mu(r)$ are given by
\begin{align}
& u(r) =  \frac{2\alpha^2+\kappa (1-\cosh r) + 2\alpha\kappa \sinh r  }{\lb 2 +\kappa +\kappa\cosh r\rb} \label{2.21}\\
&\mu(r)= \frac{1}{2D\lb 2 +\kappa +\kappa\cosh r\rb}\label{2.22}
\end{align}

We have used the large $s$ limit \cite{kal,solo4} $s\sim s+1 \sim \tilde{s} = (s+ \frac{1}{2})$, hence terms independent of $\tilde{s}$ drop out in Eq.\eqref{2.21}, also  an additional constant has been added to the potential for convenience. It is noted that the presence of the sine hyperbolic creates a metastable minimum, however, in the absence of the magnetic field the potential becomes even with two degenerate minima as shown in Fig.\eqref{interact}. 
 
\section{phase transition of the escape rate }
In this section we study the phase transition of our system in the absence of a staggered magnetic field $\alpha= 0$, as well as the phase diagram in the presence of a staggered magnetic field  $\alpha\neq 0$. For a coordinate dependent massive particle,  the existence of first-order phase transition has been shown\cite{solo5,solo6} to be determined from the condition
\begin{align}
& \bigg[U^{\prime\prime\prime}(r_b)\lb g_1+\frac{g_2}{2}\rb+\frac{1}{8}U^{\prime\prime\prime\prime}(r_b) +\omega^2\mu^{\prime}(r_b)g_2 \label{3.1}\\& +\omega^2\mu^{\prime}(r_b) \lb g_1+\frac{g_2}{2}\rb +\frac{1}{4}\omega^2\mu^{\prime\prime}(r_b)\bigg]_{\omega=\omega_b}  < 0\nonumber
\end{align}
where  

\begin{align}
&g_1 = -\frac{\omega^2\mu^{\prime}(r_b) +U^{\prime\prime\prime}(r_b)}{4U^{\prime\prime}(r_b)} \label{3.2}\\&
\nonumber
g_2 = -\frac{ 3\mu^{\prime}(r_b)\omega^2 +U^{\prime\prime\prime}(r_b)}{4\left[ 4 \mu(r_b)\omega^2+U^{\prime\prime}(r_b)\right]}  \\&\omega_b^2 =-\frac{U^{\prime\prime}(r_b) }{\mu(r_b)}\nonumber
\end{align}
and $^\prime$ represents derivatives with respect to $r$. The subscript $b$ represents the coordinate of the sphaleron at the bottom of the well of the inverted potential, and $\omega_b$ is the frequency of oscillation at the bottom of the well of the inverted potential. By setting the first derivative of the potential to zero, we obtain that the sphaleron position is  located at  $r_b = \ln\lb\frac{1+\alpha}{1-\alpha}\rb$, and the height of the potential barrier is given by
\bea
\Delta U = 2D\tilde{s}^2 \lb 1-\alpha\rb^2
\eea
Alternatively, in terms of the free energy of the system, we have that the escape rate of a particle through a potential barrier in the semiclassical approximation is obtained by taking the Boltzmann average over tunneling probabilities \cite{affl}
\begin{equation}
\Gamma \propto \int_{U_{\text{min}}}^{U_{\text{max}}} d \mathcal{E} \mathcal{P}(\mathcal{E})e^{-\beta(\mathcal{E}-U_{\text{min}})}, \quad \beta^{-1} = T
\label{decay}
\end{equation}
where  $\mathcal{P}(\mathcal{E})$ is an imaginary time transition amplitude at an energy $\mathcal{E}$, and $U_{\text{min}}$ is the bottom  of the potential energy. The transition amplitude is defined as
\begin{equation}
\mathcal{P}(\mathcal{E})\sim e^{-S(\mathcal{E})}
\label{trans}
\end{equation}
and the Euclidean action is given by
\bea
S(\mathcal{E})=2\int_{-r(\mathcal{E})}^{r(\mathcal{E})} dr \sqrt{2\mu(r)\lb U(r)-\mathcal{E}\rb}
\label{act}
\eea
where $\pm r(\mathcal{E})$ are the turning points $(\thinspace U(\pm r(\mathcal{E}))=\mathcal{E})$ at zero magnetic field for the particle with energy $-\mathcal{E}$ in an inverted potential $-U(r)$ . The factor of $2$ in Eq.\eqref{act} corresponds to the back and forth oscillation of the period in the inverted potential.  In many cases of physical interest, this integral can be computed in the whole range of energy for any given potential in terms of complete elliptic integrals. In the limit $\mathcal{E}\rightarrow U_{\text{min}}$, its value corresponds to an instanton or bounce action. All the interesting physics of phase transition in spin systems occur when the energy is very close to the top of the potential barrier, $\mathcal{E}\rightarrow U_{\text{max}}$.  In the method of steepest decent, for small temperatures $T< \hbar \omega_0$, where $\omega_0$ is the frequency of oscillation at the minimum of the potential, Eq.\eqref{decay} is dominated by a stationary point
\begin{align}
\beta=\tau(\mathcal{E})&=-\frac{dS(\mathcal{E})}{d\mathcal{E}}= \int_{-r(\mathcal{E})}^{r(\mathcal{E}}dr\sqrt{\frac{2\mu(r)}{U(r)-\mathcal{E}}} 
\label{pp}
\end{align}
which is the period of oscillation of a particle with energy $-\mathcal{E}$ in the inverted potential $-U(x)$. In the limit $\mathcal{E}\rightarrow U_{\text{min}}$, the period  $\tau(\mathcal{E})\rightarrow \infty$, that is $T\rightarrow 0$ which corresponds to an instanton  while for $\mathcal{E}\rightarrow U_{\text{max}}$,  $\tau(\mathcal{E})\rightarrow 2\pi/\omega_b$ \cite{affl}. The escape rate, Eq.\eqref{decay} in the method of steepest decent  can also be written as \cite{chud1,cl}
\begin{equation}
\Gamma \sim  e^{-\beta F_{\text{min}}}
\label{decay1}
\end{equation}
and $F_{\text{min}}$ is the minimum of the effective free energy
\bea
F = \mathcal{E}+ \beta^{-1} S(\mathcal{E})-U_{\text{min}}
\label{freenn}
\eea
with respect to $\mathcal{E}$. 

This free energy can  be used to characterize first- and second-order phase transitions in analogy with Landau's theory of phase transition, only if one can find the expression of the action $S(\mathcal{E})$ for a given mass and potential.
\subsection{Analyses with zero staggered magnetic field}
At zero staggered magnetic field, it is well-known that the ground state energy splitting of the quantum spin Hamiltonian is proportional to $J^{2s}$ which has been obtained by different methods\cite{ams,kim2,bab1,bab2}. In this section we will show how this result can be recovered from the present  formalism. At zero staggered magnetic field the effective potential, Eq.\eqref{2.21} is of the form
\begin{align}
& U(r) = \frac{ 2D\kappa s(s+1)(1-\cosh r)  }{\lb 2 +\kappa +\kappa\cosh r\rb} \label{pot2}\end{align}
Since $s\gg1$, we can approximate $s(s+1)$ as $s^2$.
In this case one can obtain the exact expression for the action, Eq.\eqref{act} in the whole range of energy by making the substitution $y= \cosh\lb\frac{r}{2}\rb$. The action becomes
\begin{align}
S(\mathcal{E})&=4s\sqrt{2(a+b)\kappa}\int_{1}^{1/\lambda} dy \frac{1}{(1+\kappa y^2)}\sqrt{\frac{1-\lambda^2y^2}{y^2-1}}
\label{act1}
\end{align}
where $\lambda^2 = \frac{2b}{a+b}$, $a=1-(2+\kappa)\mathcal{E}^{\prime}$, $b=1+\kappa \mathcal{E}^{\prime}$, and $\mathcal{E}^{\prime}= \mathcal{E}/2Ds^2\kappa$. 

Introducing the variable
\bea
x^2 = \frac{1-1/y^2}{\lambda^{\prime 2}}, \quad \lambda^{\prime 2}=1-\lambda^{2}= \frac{a-b}{a+b}
\eea
Eq.\eqref{act1} becomes
\begin{align}
S(\mathcal{\mathcal{E}})& =4s\sqrt{2(a+b)\kappa}[\mathcal{K}(\lambda^{\prime})-(1-\gamma^2) \Pi(\gamma^2,\lambda^{\prime})]
\label{act2}
\end{align}
where $\gamma^2=\lambda^{\prime 2}/(1+\kappa)$. $\mathcal{K}(\lambda^{\prime})$ and $\Pi(\gamma^2,\lambda^{\prime})$ are the complete elliptic integral of first and third kinds. Near the bottom of the potential the action is

\begin{align}
S(\mathcal{E})&\approx S(U_{\text{min}}) = 8s\arctanh(\gamma)= 4s\ln\lb\frac{\sqrt{1+\kappa}+1}{\sqrt{1+\kappa}-1}\rb\label{actt2}\end{align}
In the perturbative limit $\kappa\ll1$, Eq.\eqref{actt2} simplifies to
\begin{align} 
S(U_{\text{min}})&\approx 4s\ln\lb\frac{4}{\kappa}\rb =4s\ln\lb\frac{4D}{J}\rb
\end{align}

The ground state energy splitting in the perturbative limit  is obtained as
\bea
\Delta \mathcal{E}_0 = 2\mathscr{D}\exp\lb-\frac{S(U_{\text{min}})}{2}\rb =2\mathscr{D} \lb\frac{J}{4D}\rb^{2s}
\label{3.16}
\eea
\begin{figure}[ht]
\centering
\subfigure[ ]{%
\includegraphics[scale=0.35]{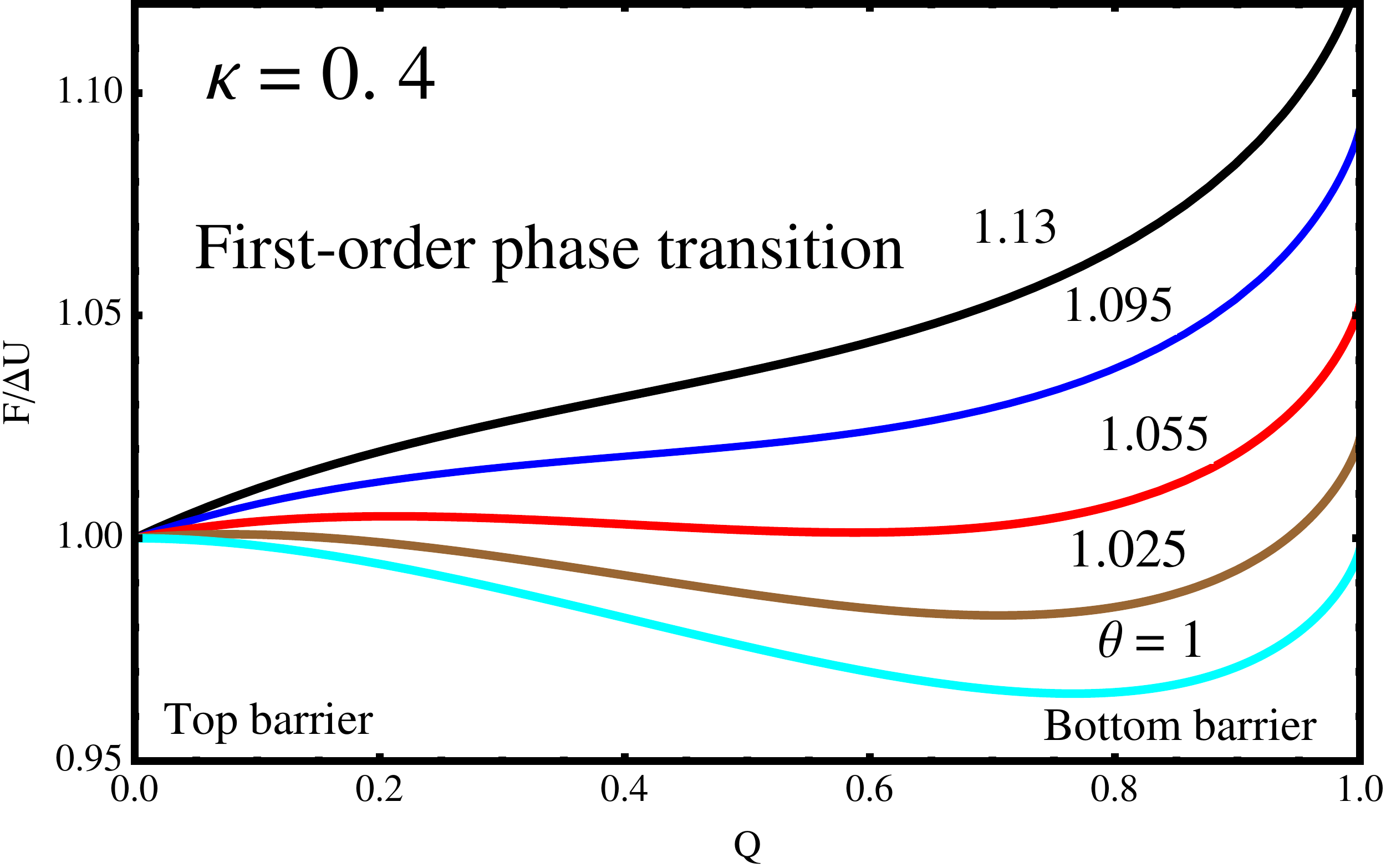}
\label{f_en}}
\subfigure[]{%
\includegraphics[scale=0.35]{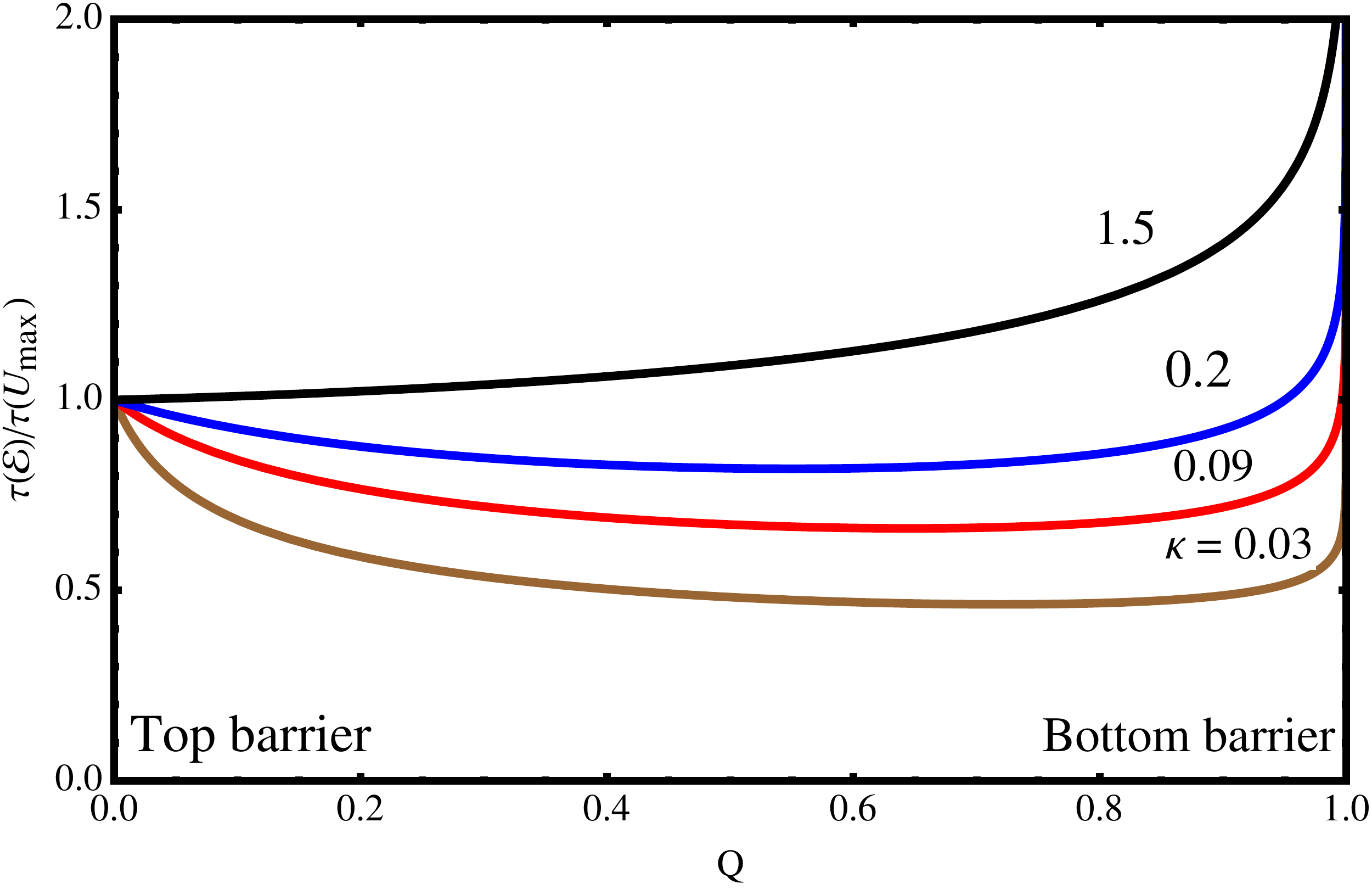}
\label{pe}}
\caption{ (a): The effective free energy of the escape rate vs Q for $\kappa=0.1$ and several values of $\theta = T/T^{(2)}_0$. (b) : The period of oscillation vs Q for several values of  $\kappa$, first-order transition. }
\end{figure}

where $\mathscr{D}$ is a prefactor which is not crucial in the present analysis.  The factor $J^{2s}$ indicates that the two classical antiferromagnetic ground state configurations are linked to each other in the $2s^{\text{th}}$ order in perturbation theory. Thus the zero magnetic field quantum spin Hamiltonian at $2s^{\text{th}}$ order can be written effectively as
\bea
\hat H \psi_{\pm}=\mathcal{E}_{\pm}\psi_{\pm}
\label{3.17}
\eea
where
\begin{align}
\psi_{\pm}&= \frac{1}{\sqrt{2}}\lb\ket{\uparrow, \downarrow}\pm\ket{\downarrow,\uparrow}\rb, \quad \Delta \mathcal{E}_0=\mathcal{E}_{+}-\mathcal{E}_{-}
\label{3.18}
\end{align}
 Thus, the ground and the first excited states are entangled states. The antisymmetric and symmetric linear superpositions are the ground and the first excited states respectively for half-odd integer spins\cite{bab,ams} while the roles are reversed for integer spins. It is noted that Kramers degeneracy does not apply in this model since we have an even number of spins.

Having obtained the  action
for all possible values of the energy, that is Eq.\eqref{act2}, the free energy Eq.\eqref{freenn} can now be written down exactly.
\begin{figure}[ht]
\centering
\includegraphics[scale=0.35]{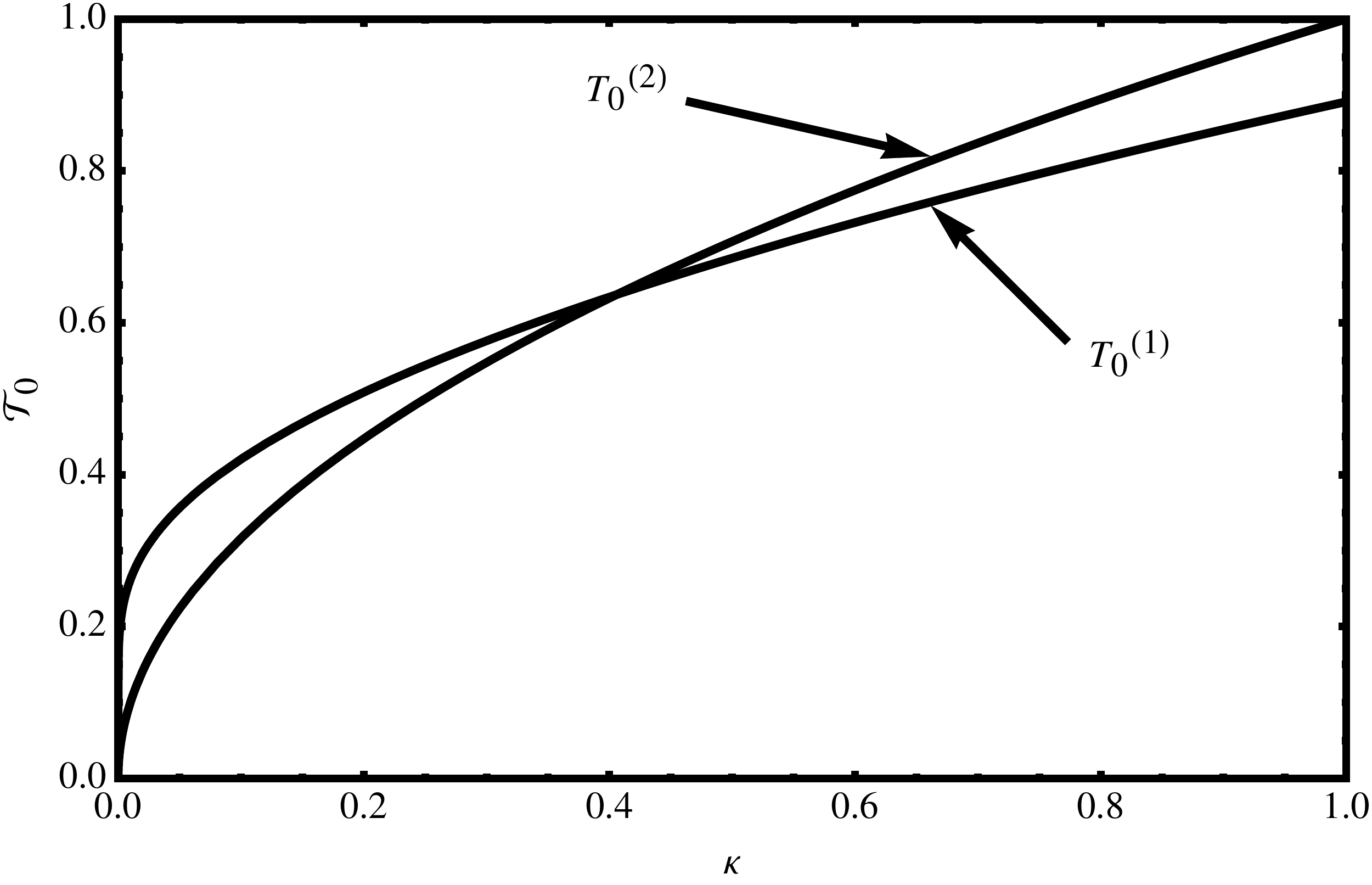}
\caption{Zero magnetic field  crossover temperatures plotted against $\kappa$. The functions increase rapidly as $\kappa$ varies between $0$ and $1$. $\mathcal{T}_0=\pi T^{(1,2)}_0/Ds$}
\label{first_order}
\end{figure}
In terms of the dimensionless energy quantity $Q= (U_{\text{max}}-\mathcal{E})/(U_{\text{max}}-U_{\text{min}})$ where $U_{\text{max}}(U_{\text{min}})$ correspond to the top (bottom) of the potential, $Q\rightarrow 0(1)$ at the top (bottom) of the potential respectively. The free energy can then be written as 
\begin{align}
F/ \Delta U&= 1-Q+\frac{4}{\pi}\theta \sqrt{\kappa(\kappa+Q)} [\mathcal{K}(\lambda^{\prime})-(1-\gamma^2) \Pi(\gamma^2,\lambda^{\prime})]
\label{freennn}
\end{align}
where $\theta = T/T_{0}^{(2)}$ is a dimensionless temperature quantity, and $T^{(2)}_0=\omega_b/2\pi$.  The  modulus of the complete elliptic integrals $\lambda^{\prime}$ and the  elliptic characteristic $\gamma$ are related to $Q$ by
\begin{align}
\lambda^{\prime 2} = \frac{(1+\kappa)Q}{\kappa+Q},\quad \gamma^2=\frac{Q}{\kappa+Q}
\label{eneq}
\end{align}

We have already known that the phase transition occurs near the top of the potential barrier, so it is required that we expand this free energy close to the barrier top. Thus, near the top of the barrier $Q\rightarrow 0$, the expansion of the complete elliptic integrals up to order $Q^3$ are given by
\begin{align}
\mathcal{K}(\lambda^{\prime}) &= \frac{\pi}{2}\bigg[1+\frac{(1+\kappa)}{4\kappa}Q+\frac{(1+\kappa)(9\kappa-7)}{64\kappa^2}Q^2\nonumber\\&+\frac{(1+\kappa)(17+\kappa(25\kappa-22))}{256\kappa^3}Q^3\bigg]\\
\Pi(\gamma^2,\lambda^{\prime}) &= \frac{\pi}{2}\bigg[1+\frac{(3+\kappa)}{4\kappa}Q+\frac{\kappa(14+9\kappa)-3}{64\kappa^2}Q^2\nonumber\\&+\frac{7-\kappa(1-\kappa(25\kappa-33))}{256\kappa^3}Q^3\bigg]
\end{align}

 The full simplification of Eq.\eqref{freennn} yields
 \begin{align}
 F/ \Delta U &= 1+(\theta-1)Q+\frac{\theta}{8\kappa}(\kappa-1)Q^2 +\frac{\theta}{64\kappa^2}(3\kappa^2-2\kappa+3)Q^3  
 \label{fredd}
 \end{align} 
This free energy looks more like the Landau's free energy, which suggests that we should compare the two free energies. The Landau's free energy  has the form
\bea 
F = a\psi^2+b\psi^4 +c\psi^6
\label{landd}
\eea
Surprisingly, the coefficient of $Q$ in Eq.\eqref{fredd} is equivalent to the coefficient $a$ in Landau's free energy. It changes sign at the phase temperature $T=T_0^{(2)}$. The phase boundary between the first- and the second-order phase transitions depends on the coefficient of $Q^2$, it is equivalent  to the coefficient $b$ in Eq.\eqref{landd}. It changes sign at $\kappa=1$. Thus $\kappa<1$ indicates the regime of first-order phase transition. These conditions for the phase boundary and the first-order phase transition can also be obtained from the criterion given in Eq.\eqref{3.1} with $x_s=r_b=0$, which corresponds to the top of the potential barrier when the magnetic field $\alpha=0$. 

 In Fig.\eqref{f_en} we have shown the plot of the free energy against $Q$ for $\kappa=0.4$ (first-order transition). In the top two curves, the minimum of the free energy is at $Q=0$. As the temperature is lowered, a new minimum of the free energy is formed. For $\theta= 1.055$ or $T_0^{(1)}=1.055T_0^{(2)}$, this new minimum becomes the same as the one at $Q=0$. This corresponds to the crossover temperature from classical to quantum regimes. 
 
  The calculation of the period of oscillation $\tau(\mathcal{E})$ yields
\begin{align}
\tau(\mathcal{E})&=-\frac{dS(\mathcal{E})}{d\mathcal{E}}= \int_{-r(\mathcal{E})}^{r(\mathcal{E})}dr\sqrt{\frac{2\mu(r)}{U(r)-\mathcal{E}}}\nonumber\\&= \frac{2\sqrt{2}}{Ds\sqrt{(a+b)\kappa}}\mathcal{K}(\lambda^{\prime})=\frac{2}{Ds\sqrt{(\kappa+Q)}}\mathcal{K}(\lambda^{\prime})
\end{align}
The plot of $\tau(\mathcal{E})$ vs $Q$ is shown in Fig.\eqref{pe} for several values of $\kappa$. The period lies in the interval $2\pi/\omega_b\leq\tau(\mathcal{E})\leq \infty$ for $0\leq Q\leq 1$. The order of phase transition can be characterized by the behaviour of the period of oscillation as a function of energy. For first-order phase transition, the period of oscillation $\tau(\mathcal{E})$ is nonmonotonic function of $\mathcal{E}$ in other words $\tau(\mathcal{E})$ has a minimum at some point $\mathcal{E}_1<\Delta U$ and then rises again, while for second-order phase transition $\tau(\mathcal{E})$ is monotonically increasing with decreasing energy\cite{cl,chud6}.  Indeed for $\kappa<1$, the period is a nonmonotonic function of energy indicating the existence of first-order phase transition. For $\kappa>1$, the period is increasing with decreasing energy which indicates a second-order phase transition. The action at the bottom of the potential, which corresponds to the instanton action i.e Eq.\eqref{actt2} can now be used to estimate the first-order crossover temperature:
\bea
T_0^{(1)}= \frac{\Delta U}{S(U_{\text{min}})}= \frac{Ds}{4\arctanh(\gamma)}, \quad \gamma\approx \frac{1}{1+\kappa}
\label{ttemp}
\eea
 For the case of second-order transition, we have
\bea
T_0^{(2)}=\frac{\omega_b}{2\pi}=\frac{Ds\sqrt{\kappa}}{\pi}
\eea
In Fig.\eqref{first_order} we have shown the plot of $T_0^{(1)}$ and $T_0^{(2)}$  against $\kappa$. The functions increase rapidly with an increase in $\kappa$ and    coincide at $\kappa=0$ and $\kappa=0.4$. At $\kappa=0.4$, we obtain $T_0^{(1)}=1.002T_0^{(2)}$. Thus, Eq.\eqref{ttemp} underestimates the crossover temperature found in Fig.\eqref{f_en}. As in the uniaxial ferromagnetic spin model \cite{cl}, one expects that both temperatures coincide for very small values of $\kappa$.  With the use of experimental parameters: $s=9/2$, $D=0.75 K$, and $J=0.12 K$ we obtain $T_0^{(1)}=0.51K$ and $T_0^{(2)}=0.43K$.
\subsection{Analyses with a staggered magnetic field}

\begin{figure}[ht]
\centering
\includegraphics[scale=0.35]{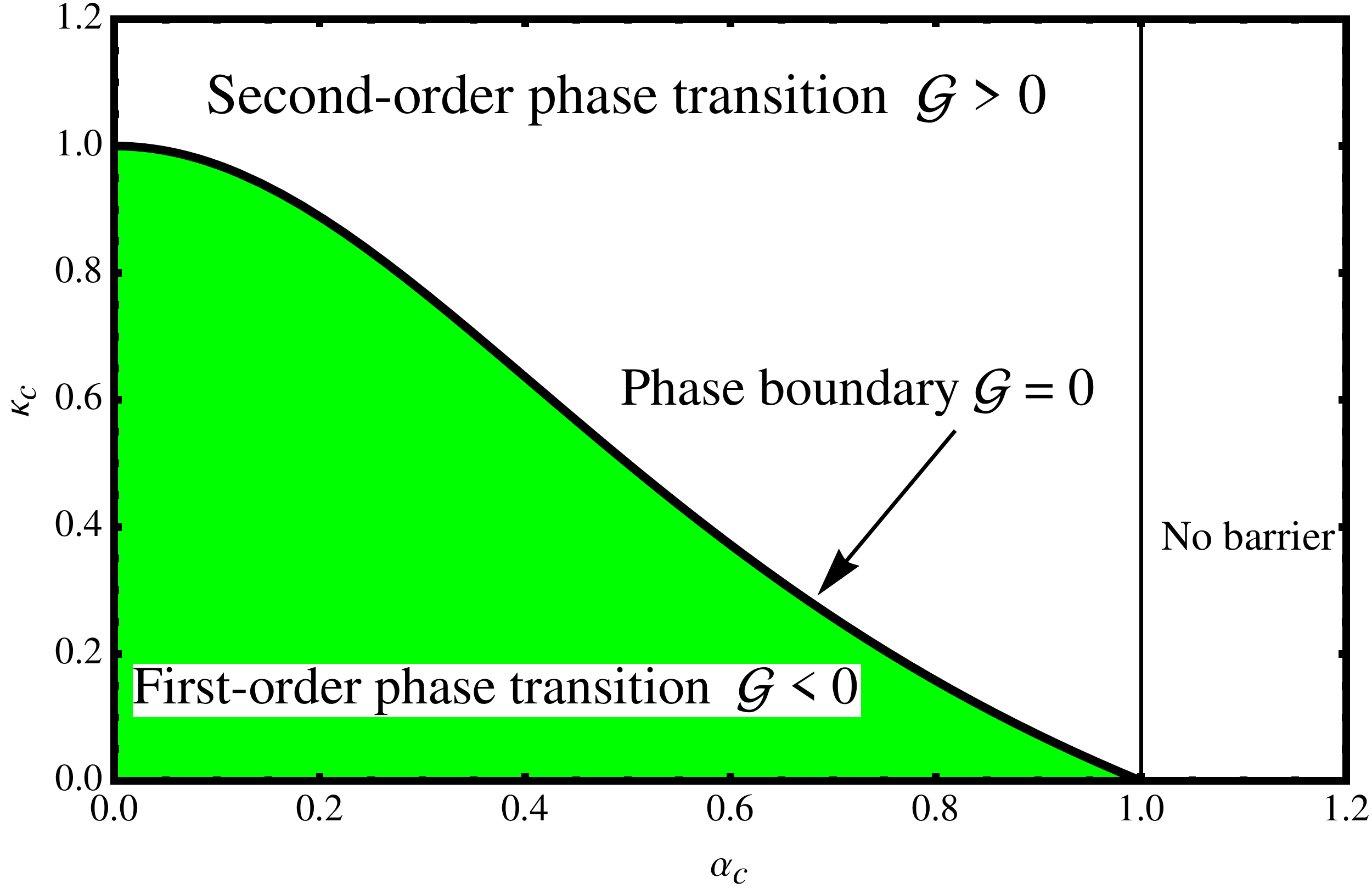}
\caption{Colour online: The phase diagram of $\kappa_c$ vs $\alpha_c$. There is no tunnelling at $\kappa_c=0$ since the individual spins $\hat{S}_{A,z}$ and  $\hat{S}_{B,z}$ commute with the Hamiltonian leading to a constant potential.}
\label{phase}
\end{figure}

In the presence of a staggered field, we would like to obtain the free energy in the whole range of energy, but this calculation is too cumbersome. So we will first use the criterion in Eq.\eqref{3.1}.  After a tedious but straightforward calculation of the derivatives in Eqs.\eqref{3.1} and \eqref{3.2}, we obtain the condition for the first-order phase transition as
\begin{align}
&\frac{ D\kappa \tilde{s}^2 (1-\alpha^2) (-1 + \kappa + \alpha^2 (1 + 2\kappa))}{8(1-\alpha^2 +\kappa)^2}<0
\end{align}
Setting this expression to zero, we obtain the phase boundary between the first- and second-order transitions as
\bea
\alpha_c = \pm \sqrt{\frac{ 1-\kappa_c}{ 1+2\kappa_c}}
\label{3.4}
\eea

where the subscript c represents the critical value of the parameters at the phase boundary. We will take only the positive sign in this expression. At zero staggered magnetic field, we obviously recover the results of the previous section. In Fig.\eqref{phase} we have shown the plot of the function $\kappa_c$ against $\alpha_c$. The plot shows a  decreasing function with increasing $\alpha_c$, at $\kappa_c=0$ we have $ \alpha_c =1$ which gives no tunnelling since the individual $z$-components of the spins commute with the Hamiltonian, thus Eq.\eqref{2.2} is an exact eigenstate which leads to  a constant potential. The shaded and unshaded regions correspond to the two regions of first- and the second-order transitions respectively, separated by a phase boundary. 
 Using the set of parameters in a realistic model of  [Mn$_{4}$]$_{2}$ dimer\cite{da,aff} $J=0.12K$, $D=0.75K$  $\Rightarrow\kappa_c=0.16$, we obtain $\alpha_c=0.80$. In the present analysis these values obviously fall in the regime of the first-order phase transition.

In order to show the analogy of these transitions to Landau's theory of phase transition as we did in the previous section,  let us consider an alternative method for deriving the critical condition Eq.\eqref{3.4}. Since we cannot compute the imaginary time action in Eq.\eqref{act} exactly, we will expand it near the top of the barrier, that is $Q\to 0$.
 
\begin{figure}[ht]
\centering
\includegraphics[scale=0.35]{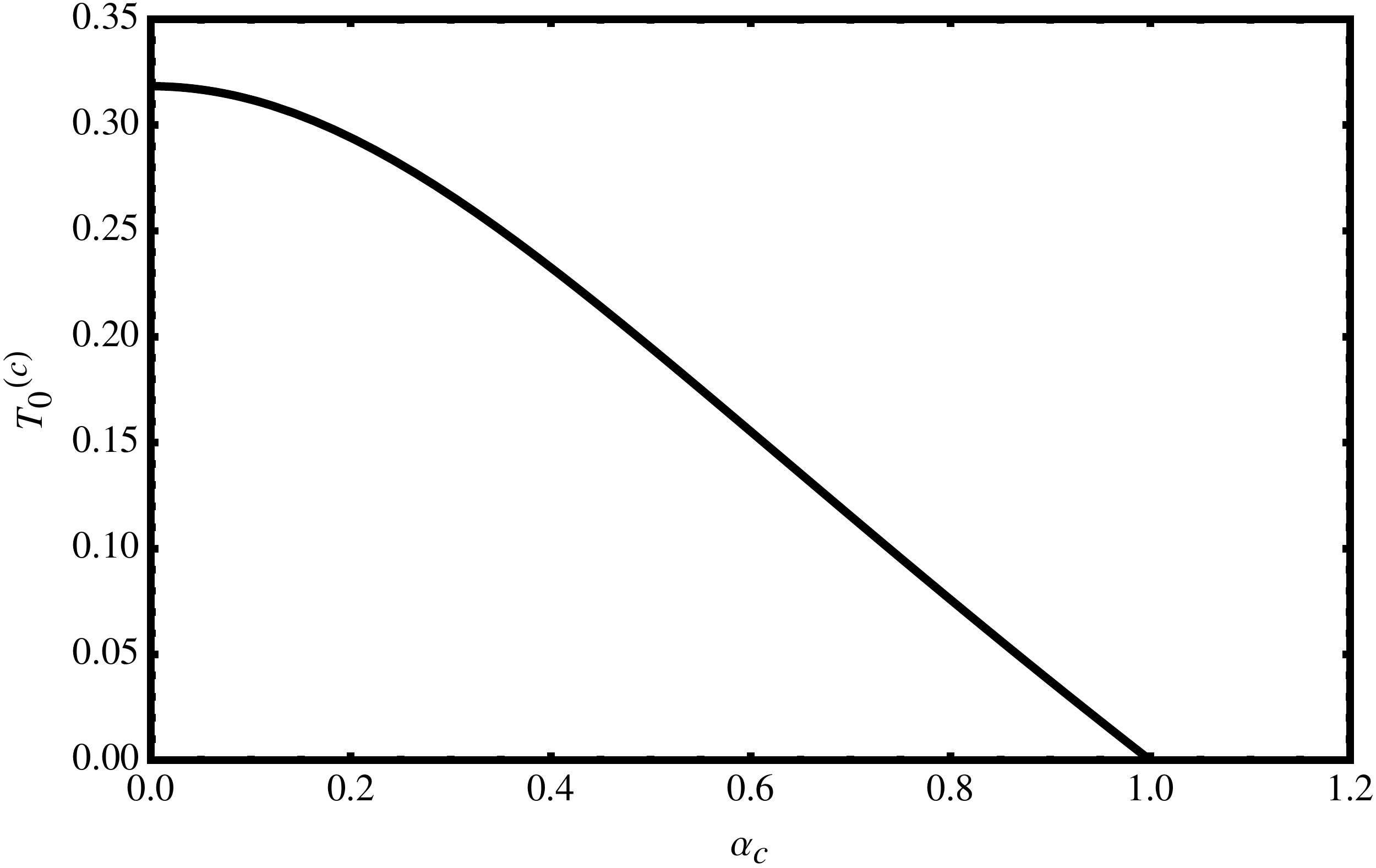}
\caption{The crossover temperature at the phase boundary between first- and second-order transitions plotted against $\alpha_c$.  $T_0^{(c)}$ has been rescaled as $T_0^{(c)}/ D\tilde{s}$. }
\label{cross}
\end{figure}
  The expansion of the imaginary time action around $r_b$ gives \cite{solo4} 
\bea
S(\mathcal{E})= \pi\sqrt{\frac{2\mu(r_b)}{U^{\prime\prime}(r_b)}}\Delta U[Q + \mathcal{G} Q^2+ O(Q^3)]
\eea
where
\begin{align}
\mathcal{G}&= \frac{\Delta U}{16U^{\prime\prime}}\bigg[\frac{12U^{\prime\prime\prime\prime} U^{\prime\prime}  + 15(U^{\prime\prime\prime })^2 }{2(U^{\prime\prime })^2}+3\lb \frac{\mu^{\prime}}{\mu}\rb \lb \frac{U^{\prime\prime\prime}}{U^{\prime\prime}}\rb \label{3.8}\\& +\lb \frac{\mu^{\prime\prime}}{\mu}\rb-\frac{1}{2}\lb \frac{\mu^{\prime}}{\mu}\rb^2\bigg]_{r=r_b}\nonumber
\end{align}
$U^{\prime\prime}(r_b)=- D\tilde{s}^2u^{\prime\prime}(r_b)/2!$, $U^{\prime\prime\prime}(r_b)=  D\tilde{s}^2u^{\prime\prime\prime}(r_b)/3!$, and $U^{\prime\prime\prime\prime}(r_b)= D\tilde{s}^2u^{\prime\prime\prime\prime}(r_b)/4!$.

 By the analogy with the Landau theory of phase transition, the phase boundary between the first- and second-order transition (see Fig.(1)) is obtained by setting the coefficient  of $Q^2$ to zero i.e $b=\mathcal{G}=0$. Using Eqns.\eqref{2.21} and \eqref{2.22} we obtain that this condition yields
\begin{align}
&\frac{ (-1 + \kappa + \alpha^2 (1 + 2\kappa))}{8\kappa(1+\alpha)^2}=0
\end{align} which again recovers Eq.\eqref{3.4} and the exact coefficient of $Q^2$ in Eq.\eqref{fredd} at $\alpha=0$. In the case of second-order transition the crossover temperature is estimated as $T_{0}^{(2)} = \omega_b/2\pi$. Using this expression and Eq.\eqref{3.4}   we obtain the crossover temperature at the phase boundary as
 \bea
T_{0}^{(c)}= \frac{ D\tilde{s}}{\pi}\frac{(1-\alpha_c^2)}{\sqrt{1+2\alpha_c^2}}=\frac{D \tilde{s}\kappa_c}{\pi}\lb\frac{ 3 }{1+2\kappa_c}\rb^{\frac{1}{2}}
\label{3.12}
\eea

The plot of $T_{0}^{(c)}$ vs $\alpha_c$(using Eq.\eqref{3.4})  is shown in Fig.(2), with the parameters for the experimental data in [Mn$_{4}$]$_{2}$ dimer \cite{da,aff}, $s=9/2$, $D=0.75K$, $\kappa_c=0.16 \Rightarrow \alpha_c=0.80$, we find $T_{0}^{(c)} =0.29K$. This crossover temperature is completely accessible as it has been experimentally demonstrated that there exist a crossover temperature below which quantum tunnelling is dominant\cite{wern4}.

  \section{ Conclusions}
In conclusion, we have investigated an effective Hamiltonian  of a dimeric molecular nanomagnet which interacts ferromagnetically or antiferromagnetically. Using the method of mapping a spin system to a particle in an effective potential, we showed that this model can be mapped to a relative coordinate dependent massive particle in a potential field. We showed that the boundary between the first-and second-order phase transitions is greatly influenced by the staggered magnetic field. The parameter values for  molecular [Mn$_{4}$]$_2$ dimer in recent experiments was shown to fall in the regime of first-order phase transition. The results obtained here are experimentally accessible.

\section{ Acknowledgments}
We thank  NSERC of Canada for financial support.

\end{document}